# Two-scale structure for giant field enhancement: combination of Rayleigh anomaly and colloidal plasmonic resonance

Mahsa Darvishzadeh-Varcheie,[1] William J. Thrift,[2] Mohammad Kamandi,[1] Regina Ragan,[2] and Filippo Capolino,[1,*]

[1]Department of Electrical Engineering and Computer Science, University of California, Irvine, CA, 92697, USA
[2]Department of Chemical Engineering and Material Science, University of California, Irvine, CA, 92697, USA
*Corresponding author: f.capolino@uci.edu

We demonstrate theoretically and experimentally a two-scale architecture able to achieve giant field enhancement by simultaneously exploiting both the Rayleigh anomaly and localized surface plasmon resonance. Metallic oligomers composed of colloidal nanospheres are well-known for the ability to strongly enhance the near-field at their plasmonic resonance. However, due to intrinsic nonlocality of the dielectric response of the metals along with their inherent loss, the achievable field enhancement has an ultimate constraint. In this paper we demonstrate that combining plasmonic resonance enhancements from oligomers, with feature size of tens of nanometers, with a Rayleigh anomaly caused by a 1-D set of periodic nanorods, having a period on the order of the excitation wavelength, provides a mean to produce enhancement beyond that constrained by losses in near field resonances. Metallic oligomers are chemically assembled in between the periodic set of nanorods that are fabricated using lithographic methods. The nanorod periodicity is investigated to induce the Rayleigh anomaly at the oligomers plasmonic resonance wavelength to further enhance the field in the oligomers' hot spots. A thorough study of this structure is carried out by using an effective analytical-numerical model which is also compared to full-wave simulation results. Experimental results comparing enhancements in surface enhanced Raman scattering measurements with and without nanorods demonstrate the effectiveness of a Rayleigh anomaly in boosting the field enhancement. The proposed structure is expected to be beneficial for many applications ranging from medical diagnostics to sensors and solar cells.

## 1. INTRODUCTION

Enhancing light-matter interaction has attracted tremendous research interest due to the reliance of various applications ranging from sensing [1–4] and solar cells [5,6] to harmonic generation enhancement [7–9] and perfect lensing [10] on local field enhancement. Owing to this fact, metallic nanoparticles have been the focus of attention for their ability to provide giant field localization at subwavelength scale thanks to the role of surface plasmon polaritons (SPPs). Particularly, nanosphere-based clusters with nanometer(nm)-sized separation have been used to achieve extreme field enhancement between metallic nanoparticles [11]. The resonance of these structures can be easily tuned by size, shape and the material of the particles. Indeed, the hotspot of the field enhancement occurs near sharp corners and tips or between nanometer-sized gaps of nanoparticles where the intensity of light is enhanced by orders of magnitude [12]. In order to attain nanoscale gap between the nanoparticles in the cluster, chemical assembly techniques have been exploited [13,14]. For example, in [15] electrohydrodynamic flow and chemical crosslinking are combined to yield gold nanospheres cluster with 0.9 nm sized gaps. Although reducing the gap size boosts the field, the field enhancement that can be achieved using subnanometer-sized gap plasmonic particles has an ultimate constraint [16,17]. Strictly speaking, the intrinsic nonlocality of the dielectric response of the metal causes this ultimate constraint rather than its inherent losses. From a physical point of view, the charge density is not perfectly restrained on the surface but is slightly diffused into the volume of the nanoparticle [18] which restricts the field enhancement. Realization of sensitive sensors requires even higher levels of field enhancement than what is currently achieved in current plasmonic chemically assembled surfaces. Consequently, it is of crucial importance to make use of other pathways along with plasmonic resonances to further enhance the field.

Periodic structures have been extensively utilized to tailor light-matter interaction. Specifically, one promising way of further enhancing the field and overcoming the aforementioned limit is to employ the constructive interference of scattered fields in periodic arrays of metallic nanostructures. This interesting phenomenon, also known as Rayleigh anomaly, leads to sharp resonant-like peaks in the scattering, absorption and emission spectra at the wavelength close to the period of the structure. With this goal of combining the capability of plasmonic resonances with Rayleigh anomaly, periodic arrays of metallic nanoparticles have been used to obtain remarkably narrow plasmonic resonance [19–23] in which the coupling between local surface plasmon resonance (LSPR) modes of the metallic nanoparticles and the lattice modes of the array has further enhanced the field compared to the field enhancement taken of individual metallic nanoparticles. However, it is extremely difficult to achieve giant field enhancement with this design since periodicity of these structures needs to be precise and demands using lithography techniques which restrict the size and the shape of individual nanoparticles.

In this paper, we introduce a novel architectural scheme to achieve giant field enhancement by combining the Rayleigh anomaly and plasmonic resonance of nanoparticles. Compared to previous studies, we use two different sets of structures concurrently to realize high localization of fields. We refer to it as a "two-scale" method since the periodic array of scatterers and the chemically assembled oligomers have completely different scales and fabrication methods. Particularly, we use a periodic set of metallic nanorods deposited over a substrate along with chemically assembled metallic oligomers between the nanorods (Fig. 1). The period of the nanorods is chosen such that its scattering resonance due to Rayleigh anomaly overlaps with the LSPR mode of the oligomers. In such a design, when the structure is illuminated with an incident wave, the coherent scattering from the nanorods (Rayleigh anomaly) boosts the field and transfers energy to the LSPR mode of the oligomer nanoantennas which leads to strong field enhancement. The deposition of the nanorods can be done using lithography or by using near-field electrospinning (NFES) method [24]. After depositing the nanorods with a suitable period, self-assembly techniques can be used to form cluster of oligomers with nm-sized gap. Notice that, in our proposed structure, we separate the fabrication of the periodic array from the formation of clusters of oligomers (which is done by self-assembly). Thus, oligomers which provide the field localization in their plasmonic resonance are not restricted for their shape or size. We experimentally verify the ability of our structure for field enhancement by exploiting it for surface-enhanced Raman spectroscopy (SERS) which is a vibrational spectroscopy technique for identification of trace amounts of analytes. Approximately, SERS is proportional to the quadratic power of the field enhancement, if we assume that the excitation and scattering frequency are close to each other. Therefore, probing SERS enhancement is a suitable tool to reveal the usefulness of the two-scale structure.

The rest of the paper is organized as follows: first, we analytically demonstrate the effect of periodicity on the resonance wavelength of individual nanorods. Then we investigate the efficacy of the two-scale structure by tuning the nanorods period on boosting the near-field enhancement in hot-spot of oligomers. In the end, we illustrate the experimental SERS result and compare it with our defined figure of merit based on the field enhancement results achieved by full-wave simulations. Finally, we will conclude the paper with some remarks.

## 2. STATEMENT OF THE PROBLEM

Our proposed two-scale structure consists of oligomers of nanospheres with subwavelength diameter and gaps, in the middle of a one dimensional (1-D) periodic array of plasmonic nanorods with a period close to the resonance wavelength of the oligomers as shown in Fig. 1(a). As it is well established, surfaces of plasmonic nanospheres in oligomers provide hot-spots for large electric field due to their plasmonic resonance. As already mentioned, the goal of this paper is to combine the plasmonic resonance of the oligomers along with the Rayleigh anomaly due to the presence of the periodic 1-D nanorods to achieve further field enhancement compared to oligomers alone when illuminated with an incident plane wave. It is worth mentioning that the resonance frequency of oligomers depends on the number, size and the gap between the nanospheres which form the oligomers [3]. On the other hand, the wavelength at which the Rayleigh anomaly occurs depends on the period of the structure, host medium and angle of incidence. For the purpose of evaluating the ability of the nanostructure in boosting the electric field, we define the electric field enhancement ($FE$) as a figure of merit [3,11,15,25]

$$FE = \frac{|\mathbf{E}^{\text{tot}}(\mathbf{r})|}{|\mathbf{E}^{\text{inc}}(\mathbf{r})|} \quad (1)$$

where $|\mathbf{E}^{\text{tot}}(\mathbf{r})|$ is the magnitude of the total electric field at location $\mathbf{r}$, and $|\mathbf{E}^{\text{inc}}(\mathbf{r})|$ is the magnitude of the incident electric field at the same location in the absence of the structure. The electric $FE$ indicates the ability of a nanoantenna system to enhance the electric field locally with respect to the incident illuminating field. Since the sizes of the 1-D periodic nanorods and the oligomers are incomparable, for our proposed structure, with good approximation we can neglect the coupling from the oligomers to the nanorods. Thus, the *total* field enhancement achieved by the two-scale structure ($FE_t$) can be approximate with the product of field enhancement obtained by periodic nanorods ($FE_r$) and the oligomers ($FE_o$) individually:

$$FE_t \approx FE_r \times FE_o \quad (2)$$

The mechanisms responsible for these two field enhancements are distinct: in oligomers, it is the plasmonic resonance whereas in the 1-D periodic array of nanorods, the field enhancement is due to the Rayleigh anomaly (in [26] it is referred to as structural resonance).

In the two-scale structure, as it is shown in Fig. 1(a), the colloidal oligomers consist of an arbitrary number of plasmonic nanospheres with diameter "$d_s$", gap "$g$" and relative electric permittivity "$\varepsilon_s$". For what matters in this paper when their density is not high, these subwavelength oligomers can be considered as isolated scatterers by neglecting their mutual electromagnetic couplings. The larger scale is made of rods with diameter "$d_r$" and relative permittivity "$\varepsilon_r$", have period "$p_r$" along the *x*-direction and are long compared to the wavelength. The rods are placed on top of a substrate with permittivity "$\varepsilon_{\text{sub}}$" under a medium with relative permittivity "$\varepsilon_h$".

In this paper, the monochromatic time-harmonic convention $\exp(-i\omega t)$ is implicitly assumed and its notation is suppressed in the following. In all equations, bold fonts are used for vector quantities in phasor domain, and a bar under a bold font is used for dyadic quantities. Unit vectors are bold with a hat on top.

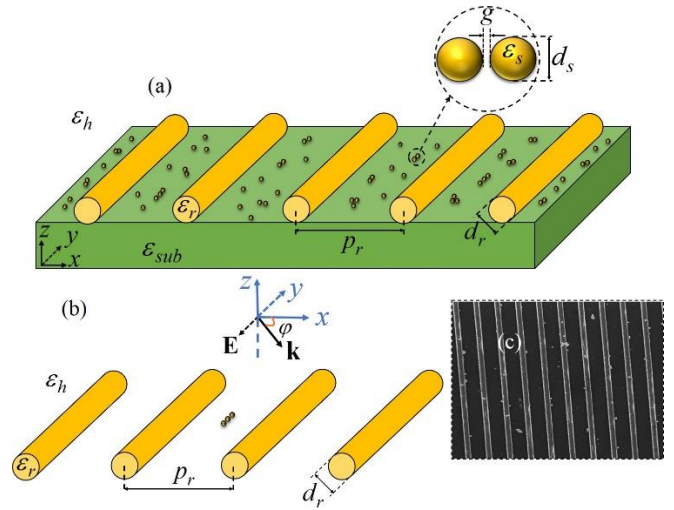

Fig. 1. (a) Proposed "two-scale" structure for enhancing the electric field at optical frequencies. It is made of oligomers of colloidal plasmonic nanospheres assembled within the nanorods of a 1-D periodic array, on top of a substrate. (b) Simplified two-scale structure for analytical investigation consisting of plasmonic trimers in the middle of the 1-D periodic array of plasmonic nanorods in a homogeneous host medium. (c) SEM image of the fabricated structure, which consists of gold oligomers placed between periodic gold nanorods, on the glass substrate.

## 3. ANALYTICAL MODEL

We present an analytical approach for obtaining the field enhancement of our proposed structure. While the method is general, in order to simplify the analytic investigation of the two-scale structure, we limit our study to linear oligomers of plasmonic nanospheres (dimers or trimers) that are placed in between plasmonic nanorods inside a homogeneous host medium with permittivity $\varepsilon_h$, as shown in Fig. 1(b). We consider that the nanorods are periodic along the *x*-axis and are long enough (compared to the wavelength, along the *y*-axis) so in absence of the nanospheres, we presume $\partial/\partial y = 0$ for any field quantity. Furthermore, as mentioned earlier, we consider the oligomers in between the nanorods to be subwavelength and separated from each other with enough distance. Therefore, we neglect not only the coupling among oligomers but also the effect of coupling of oligomers on the nanorods but not vice-versa, i.e. we will focus on the field scattered by the nanorods that excites the oligomers. As previously shown in Eq. (2), the total field enhancement is the multiplication of the field enhancement caused by the 1-D periodic array of nanorods i.e. $FE_r$ and the field enhancement produced due to the plasmonic resonance of the oligomers i.e. $FE_o$ since the mechanism of each type of field enhancement is distinct. We start by obtaining the scattered field from the nanorods and then investigate the field enhancement caused by the oligomers in the next sections, respectively.

### A. Enhanced Scattered Electric Field by Nanorods

The goal of this section is to calculate the *external* electric field at the oligomers location, which is the summation of the incident field and the scattered field from the nanorods. We start our analysis by assuming an oblique incident plane wave with transverse electric (TE) polarization along $y$, as noted in Fig. 1(b), as $\mathbf{E}^{\text{inc}}(\mathbf{r}) = \hat{\mathbf{y}} E_0 e^{ik_x x} e^{-ik_z z}$, where $k_x = k_h \cos(\varphi)$, $k_z = k_h \sin(\varphi)$, $k_h = k_0 \sqrt{\varepsilon_h}$ is the wavenumber of host medium $k_0 = \omega\sqrt{\mu_0 \varepsilon_0}$ is the wavenumber of free space and $\varphi$ is the angle of incidence with respect to $x$-axis. Since the incident field is polarized along the nanorods axis (the $y$-direction), we model each nanorod with an equivalent electric current $I_n$ along the nanorod. The induced current of the $n^{th}$ nanorod at location $\mathbf{r}_n$ ($n^{th}$ nanorod's location) is given by

$$I_n = \alpha_r E_r^{\text{loc}}(\mathbf{r}_n) \qquad (3)$$

where $E_r^{\text{loc}}(\mathbf{r}_n)$ is the local electric field at the $n^{th}$ nanorod location polarized in $y$-direction and $\alpha_r$ is the electric polarizability of an infinitely long rod with diameter $d_r$ (the cross section area is $A = \pi d_r^2 / 4$) defined as [27,28]

$$\alpha_r = \frac{-i\omega\varepsilon_0(\varepsilon_r - \varepsilon_h)A}{1 - i\frac{k_h^2}{4}A(\varepsilon_r - \varepsilon_h)} \qquad (4)$$

Note that since the currents and electric field are all along the $y$-axis, we can simplify all the equations to scalar equations. The rods are periodic in the $x$-direction with period of $p_r$, and the location of the $n^{th}$ nanorod is $\mathbf{r}_n = \mathbf{r}_0 + np_r\hat{\mathbf{x}}$ where $\mathbf{r}_0$ denotes the reference rod location which we assume to be $\mathbf{r}_0 = 0\hat{\mathbf{x}} + 0\hat{\mathbf{z}}$. Therefore, the corresponding location of the $n^{th}$ nanorod is $\mathbf{r}_n = np_r\hat{\mathbf{x}}$. The current on the $n^{th}$ nanorod is then $I_n = I_0 e^{ik_x np_r}$, where $I_0$ is the reference current (of the 0$^{th}$ nanorod) [29] Thus, the external electric field at a general location $\mathbf{r}$ is written as the summation of the incident electric field and the scattered field from the 1-D periodic set of nanorods as [30]

$$E_r^{\text{ext}}(\mathbf{r}) = E^{\text{inc}}(\mathbf{r}) + i\omega\mu I_0 G^\infty(\mathbf{r},\mathbf{r}_0) \qquad (5)$$

where $G^\infty(\mathbf{r},\mathbf{r}_0)$ is the scalar periodic Green's function of magnetic vector potential with respect to the $y$-direction. The definition of elements of Green's function is presented in the Methods (A). The nanorods' local field at the reference rod location to be used in Eq. (3) is

$$E_r^{\text{loc}}(\mathbf{r}_0) = E^{\text{inc}}(\mathbf{r}_0) + i\omega\mu_0 I_0 \breve{G}^\infty(\mathbf{r}_0,\mathbf{r}_0) . \qquad (6)$$

Note that in the definition of local field (Eq.(6)), the contribution from the reference nanorod itself should be removed [29] and the value of the "regularized" scalar periodic Green's function $\breve{G}^\infty(\mathbf{r}_0,\mathbf{r}_0)$ is determined by the limit $\breve{G}^\infty(\mathbf{r}_0,\mathbf{r}_0) = \lim_{\mathbf{r}\to\mathbf{r}_0}[G^\infty(\mathbf{r},\mathbf{r}_0) - G(\mathbf{r},\mathbf{r}_0)]$, since both $G^\infty(\mathbf{r},\mathbf{r}_0)$ and $G(\mathbf{r},\mathbf{r}_0)$ are singular at $\mathbf{r} = \mathbf{r}_0$, where $G(\mathbf{r},\mathbf{r}_0)$ is the two dimensional scalar Green's function of magnetic vector potential as defined in the Methods (A). Substituting $E_r^{\text{loc}}(\mathbf{r}_0)$ from Eq.(6) in Eq.(3), the current along the reference nanorod is found as

$$I_0 = \frac{\alpha_r E^{\text{inc}}(\mathbf{r}_0)}{1 - i\omega\mu_0\alpha_r\breve{G}^\infty(\mathbf{r}_0,\mathbf{r}_0)} . \qquad (7)$$

By substituting Eq. (7) into Eq.( 5), one can easily calculate the external electric field as a result of incident field and the multiple scattering by nanorods at an arbitrary oligomer location. One should note, as shown in Methods (B) that the periodic Green's functions in Eq. (7) and Eq. (5) have the terms $k_{zp} = \sqrt{\varepsilon_h k_0^2 - (k_x + 2\pi p / p_r)^2}$, with $p = 0, \pm 1, \pm 2,...$, at the denominator. Therefore, when a given $k_{zp} \approx 0$ one has $G^\infty(\mathbf{r},\mathbf{r}_0) \approx B/k_{zp}$ and $\breve{G}^\infty(\mathbf{r}_0,\mathbf{r}_0) \approx C/k_{zp}$. At the Rayleigh's anomaly wavelength (which will be discussed later in Eq. (12) with more details), the scalar periodic Green's function of the magnetic vector potential has a "spectral" singularity occurring when $k_{zp} = 0$ i.e., when $(k_x + 2\pi p / p_r) = \pm\sqrt{\varepsilon_h}k_0$. Thus, we can approximate the reference current in Eq. (7) as $I_0 \approx E^{\text{inc}}(\mathbf{r}_0) / (-i\omega\mu_0 \breve{G}^\infty(\mathbf{r}_0,\mathbf{r}_0))$. However, $E_r^{\text{ext}}(\mathbf{r})$ in Eq. (5), has a finite value i.e., the electric field enhancement of the nanorods evaluated in the middle of two nanorods is equal to 2 (This issue is investigated comprehensively in the Methods (B)). Hence, the incident field intensity is enhanced by a factor of 4. Indeed, in the Raman spectroscopy, when the scattered frequency is close to the incident one, the maximum SERS intensity enhancement is 16.

In the next section, using these external field calculated at the location of an arbitrary oligomer, we will find the total electric field at the hot-spot of the oligomer.

### B. Total Electric Field at Oligomers' Hot-spot

In the previous section we only investigated the field generated by the nanorods, however in the two-scale structure discussed in this paper we consider an oligomer with subwavelength gap spacing placed between two adjacent nanorods. The field calculated in the presence of nanorods in Eq.( 5) acts as the external field applied to an oligomer. We employ the single dipole approximation (SDA) method [3,31] to model each of the $M$ nanospheres of an oligomer, with a single electric dipole moment $\mathbf{p}_m$, with $m = 1, 2,..M$. Indeed, due to the subwavelength size and the material of the nanospheres, we neglect their magnetic dipole and quadrupole moments. The induced electric dipole moment of the $m^{th}$ nanosphere at location $\mathbf{r}_m$ is found as

$$\mathbf{p}_m = \alpha_s \mathbf{E}_s^{\text{loc}}(\mathbf{r}_m) \qquad (8)$$

where $\alpha_s$ is the electric polarizability of every nanosphere, assumed to be isotropic [31,32] and $\mathbf{E}_s^{\text{loc}}(\mathbf{r}_m)$ is the local electric field at the $m^{th}$ nanosphere's location. The local electric field is the summation of the external field given by Eq. (5) and the field scattered by all the other nanospheres of the oligomer. Thus, the local electric field at the $m^{th}$ nanosphere's location is given by

$$\mathbf{E}_s^{\text{loc}}(\mathbf{r}_m) = \mathbf{E}^{\text{ext}}(\mathbf{r}_m) + \sum_{\substack{l=1 \\ l\neq m}}^{M} \underline{\mathbf{G}}(\mathbf{r}_m,\mathbf{r}_l) \cdot \mathbf{p}(\mathbf{r}_l) \qquad (9)$$

where $\underline{\mathbf{G}}(\mathbf{r}_m,\mathbf{r}_l)$ is the dipole dyadic Green's function providing the electric field as defined in [3]. To find the electric dipole moment $\mathbf{p}(\mathbf{r}_l)$ of the $m^{th}$ nanosphere, we combine Eq. (8) and Eq. (9), and solve the linear system ($m = 1, 2, ...M$) [3,33]

$$\sum_{l=1}^{M} \underline{\mathbf{A}}_{ml} \cdot \mathbf{p}(\mathbf{r}_l) = \alpha_s \mathbf{E}^{ext}(\mathbf{r}_m), \quad \underline{\mathbf{A}}_{ml} = \begin{cases} \mathbf{I} & l=m \\ -\alpha_s \underline{\mathbf{G}}(\mathbf{r}_m,\mathbf{r}_l) & l \neq m \end{cases} \quad (10)$$

The total scattered field by the rods $E_r^{ext}$ in Eq. (5)), acts as the external field for the oligomers ($\mathbf{E}^{ext}$ which has nonzero component only in the $y$-direction). For simplicity, for a linear oligomer along the $y$-direction, the total electric field at the hot-spot (interparticle gap of the oligomer) along the $y$-direction reads

$$E^{tot}(\mathbf{r}_{obs}) = E_r^{ext}(\mathbf{r}_{obs}) + \frac{\sum_{m=1}^{M} \alpha_s g(\mathbf{r}_{obs},\mathbf{r}_m) E_r^{ext}(\mathbf{r}_m)}{1 - \sum_{\substack{l=1 \\ l \neq m}}^{M} \alpha_s g(\mathbf{r}_m,\mathbf{r}_l)} \quad (11)$$

where $g(\mathbf{r}_{obs},\mathbf{r}_m)$ is the $\hat{\mathbf{y}}\hat{\mathbf{y}}$ element of the dipole dyadic Green's function defined in [3]. By dividing Eq. (11) over incident electric field, we can obtain total field enhancement of Eq. (1).

So far, we developed a rigorous analytical method to calculate the field enhancement of our proposed structure. Using the above method, in the next section, we obtain the field enhancement for different cases of our proposed two-scale structure.

## 4. RESULTS AND DISCUSSION

### A. Rayleigh Anomaly

Our goal is to exploit the Rayleigh anomaly to enhance local fields beyond that which can be achieved solely using LSPR. The Rayleigh anomaly arises in a periodic array of scatteres due to the coherent interaction in multiple scattering. This interaction yield a geometric sharp "resonance-like" peak which appears when the wavelength of the incident light matches the periodicity of the structure [26,34,35]. We can achieve our goal by carefully choosing the period of an array of nanorods such that the spectral location of the anomaly is commensurate with the plasmonic resonance of oligomers. In this system, the nanorods deliver enhanced local field to the plasmonic oligomers which then enhance it further.

The Rayleigh anomaly in periodic structures becomes visible as a sudden change of a measurable parameter such as transmission or reflection from a surface under the incident wave when the wavelength or angle of incidence is varying and happens at the wavelength [36,37]

$$\lambda_r = \frac{p_r}{p}\sqrt{\varepsilon_h}\left(\pm 1 - \cos\varphi\right), \quad p = \pm 1, \pm 2,... \quad (12)$$

For normal (to the plane of periodicity) angle of incidence and assuming the host medium to be the vacuum, a Rayleigh anomaly occurs when the wavelength of incident wave matches the array period. To demonstrate this phenomenon more clearly, we consider a 1-D periodic set of gold nanorods in a homogeneous host medium (vacuum), when the structure is excited by a normal plane wave polarized along the nanorod, as shown in Fig. 2, and investigate the electric field enhancement for various nanorods diameter and periodicity. The electric field enhancement is calculated based on Eq. (1), in the middle of the unit cell of a periodic array of gold nanorods. We use the Drude model for gold permittivity of the nanorods as in [38]. We consider rods' diameters of 40 and 80nm and different array periods of 545, 633 and 785 nm.

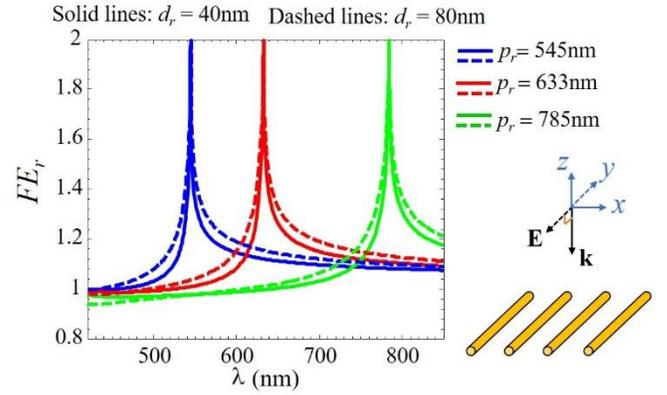

Fig. 2. Electric field enhancement achieved in the middle of the unit cell of a periodic array of gold nanorods, in a vacuum host medium, when the nanorods are excited by normal incidence plane wave polarized along the rod axis. (Field is evaluated at a point equidistant from the two adjacent rods.) Solid and dashed lines represent the result pertaining to rods' diameter of 40nm and 80 nm respectively. The period of nanorods assumed to be 545nm, 633nm and 785 nm, shown with blue, red and green respectively.

As it is clear from Fig. 2, the periodic structure possesses a sharp peak at the wavelength that is commensurate to the period of the structure when it is illuminated with a normal plane wave polarized along the nanorods. Moreover, as discussed in section (3. A) and also in Methods (B), the magnitude of the peak is equal to 2, independent of the value of the rod diameter and period. This is a striking result since for most of the spectroscopy applications, the signal enhancement is proportional to the fourth power of electric field enhancement. Therefore, spectroscopy signals can be significantly enhanced by simply using a two-scale structure. It is worth mentioning that, although the diameter of rods does not play any role in determining the resonance frequency when it is subwavelength, larger diameters provide field enhancement in wider bandwidth.

Despite the importance of normal incidence, it is interesting to study the effect of the wave incident angle on the frequency and strength of the resonance since in experiments the structure will be excited by a Gaussian beam which can be represented as a weighted summation of multiple plane waves coming at different angles. To that end, we consider a 1-D periodic array of gold nanorods with diameter of 80 nm and period of 545 nm, located in a homogeneous host medium (vacuum). We assume a plane wave polarized along the rods axis, arriving at an angle $\varphi$ with the plane of period, as shown in Fig. 3. We calculate the electric field enhancement in the middle of an array unit cell, as a function of incident angle and wavelength. Results in Fig. 3 describe that the maximum electric field enhancement is obtained at normal incidence ($\varphi = 90^o$), at the wavelength that matches the period of structure (geometric resonance at 545nm). For other angles of incidence ($\varphi \neq 90^o$), the structure provides a field that peaks at both larger and smaller wavelengths (compared to periodicity) as is also clearly illustrated in Eq. (12).

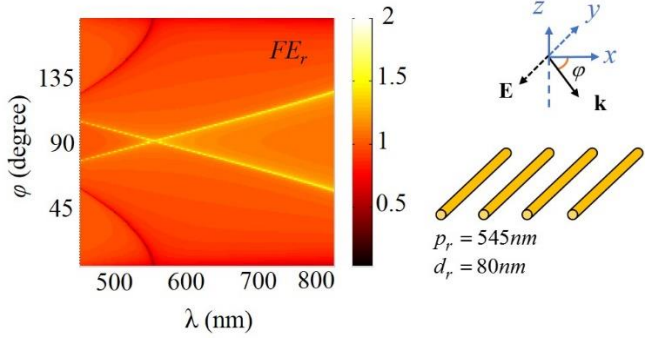

Fig. 3. Electric field enhancement $FE_r$ versus wavelength and plane wave angle of incidence. $FE_r$ is calculated in the middle of a unit cell of the array (at the same distance from the two nearest nanorods). Gold nanorods have diameter $d_r = 80$ nm and period $p_r = 545$ nm, all supposed to be in vacuum for simplicity.

### B. Two-scale structure

So far, we have only used the structural field enhancement due to the 1-D periodic set of gold nanorods. However, the ultimate goal of this paper is to combine this coherent kind of enhancement due to the larger scale of fabrication with the one caused by gold oligomers' plasmonic resonance In doing so the wavelength of geometric field enhancement (i.e., the array's period) should be tuned close to the localized surface plasmon resonance of the oligomers to achieve an even stronger electric field than that obtained by the oligomers alone [3,14]. Fig. 4, illustrates the electric field enhancement versus wavelength of the incident plane wave for a linear trimer of gold nanospheres located in the middle of a unit cell of the periodic array of gold nanorods, as shown in Fig. 1(b). The maximum electric field enhancement achieved by a linear oligomer occurs if the polarization of incident field is parallel to the oligomer axis [3,11,39]. So, we assume that the trimer axis is parallel to the nanorods and also the polarization of incident plane wave as shown in Fig. 4. We consider two different cases for the period $p_r$ of gold nanorods: (i) when the period is far from the resonance wavelength of the trimer (Fig. 4(a)) and (ii) when the period is the same as the trimer resonance wavelength (Fig. 4(b)).

In both cases, the diameter of each nanosphere of the trimer and the gap spacing between them are fixed as $d_s = 40$ nm and $g = 5$ nm respectively. If we apply the SDA method in Eq. (11) to calculate the trimer resonance in the vacuum, the resonance wavelength would be 496 nm. Therefore, for the first abovementioned case, we choose $p_r = 750$ nm (this is pretty arbitrary, and we chose this number for the period to be far from the trimer's plasmonic resonance). The zoomed-in inset in Fig 4. (a) shows the nanorods geometric resonance which happens at the wavelength close to the period. It is clear that since the 1-D periodic set of nanorods does not resonate at 496 nm, the field enhancement of the two-scale structure (rods + trimer) does not differ with the field enhancement of the trimer without periodic nanorods. It is also worth noting that the field enhancement of the two-scale structure has another small peak around the wavelength near to the period of the nanorods. This is due to the Rayleigh anomaly. Note, however, that because the plasmonic resonance of the trimer does not coincide with the Rayleigh anomaly, the total field enhancement is not further boosted in this case.

In the second case in Fig. 4(b), the period of nanorods is chosen to be the same as the trimer resonance wavelength, i.e. $p_r = 496$ nm. As Fig. 4(b) illustrates the field enhancement of the two-scale structure has only one peak. In this case the field enhancement is approximately equal to the product of electric field enhancement achieved by each of the two different structures individually as shown in Eq.( 2).

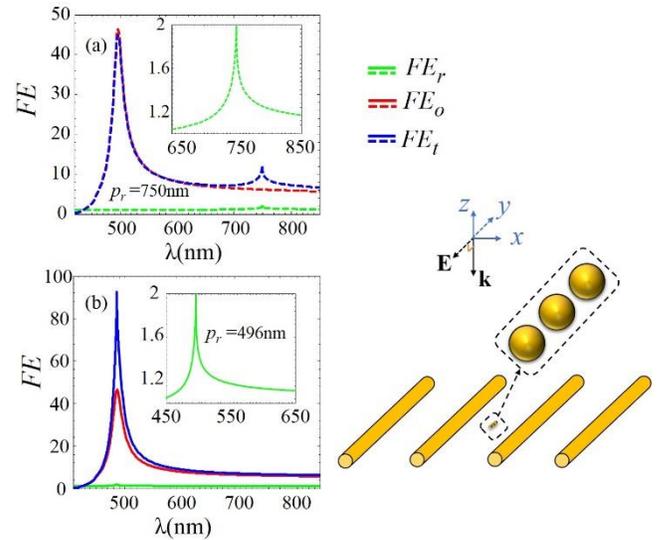

Fig. 4. Electric field enhancement versus wavelength of the incident light obtained by using the two-scale method, which consists of mixing a gold linear trimer placed in between the 1-D periodic array of gold nanorods. The nanorods' period and the trimer resonance plasmonic wavelength are (a) different, and (b) the same. The diameter of nanorods and nanospheres, as well as the gap between spheres, are kept constant as 80, 40 and 5 nm, respectively. The nanorod array period is $p_r = 750$ nm in the first case (a) and $p_r = 496$ nm in the second one (b).

So far, we have assumed plane wave as the incident field in our analysis. However, in reality, used laser beams possess Gaussian-like spatial field distribution. In order to study the ability of our proposed two-scale structure in further enhancing the field under Gaussian beam illumination, we recall that any beam can be decomposed into an infinite set of plane waves with different amplitudes and angles of incidence. Therefore, we approximate the Gaussian beam with a finite set of plane waves ranging from -30 degrees to 30 degrees from the normal incidence with different amplitudes and use our aforementioned analysis of plane waves to obtain the field in the presence of the periodic nanorods (note that this is a good approximation since the waist of the beam is much larger than the wavelength and the weights of plane waves with other angles of incidence are negligible). It is important to note that the waist of the Gaussian beam $w_0$ [40] should be chosen large (compared to the wavelength) to excite a sufficient number of nanorods (more than 15 nanorods). To that end, we excite our proposed structure using a Gaussian beam with minimum waist $w_0 = 4$ μm (see Fig 5) assuming the period of the nanorods to be $p_r = 496$ nm. The diameter of nanorods, and diameter and gap spacing of trimer are kept constant as the case studied in Fig. 4(b) as 80nm, 40 nm and 5 nm, respectively. Note that since the waist of the Gaussian beam is large compared to the wavelength, the weight of the normal plane wave and of plane waves with incidence angles close to normal incidence are larger than that relative to other angles in forming the Gaussian beam. Therefore, we do not expect the results of the field enhancement due to the Rayleigh anomaly to be drastically different from the results of normal incidence illumination. We mainly expect a broadening of the peak when varying wavelength, i.e., a wider bandwidth of field enhancement caused by the array of rods. However, as results clearly illustrate, the peak of the field enhancement for the Gaussian beam is slightly smaller than that of a plane wave in Fig. 2, since for this scenario the power of the beam is distributed over different angles. Besides, in this case, there are two peaks (see the green line) which do not happen at the period of the nanorods. To provide the reason for this behavior, note that as Fig. (3) illustrates, when the wavelength of the incident light matches the period, field enhancement peaks at normal incidence, however, for larger and smaller wavelengths, the field enhancement peaks at two oblique angles. Therefore, for wavelengths close to (and not equal to) the period of the nanorods, the field enhancement peaks.

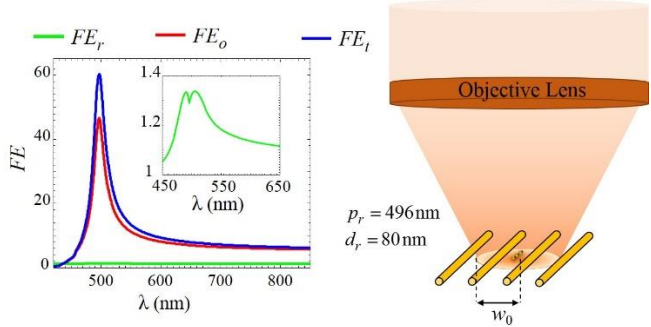

Fig. 5. Electric field enhancement *versus* wavelength of an incident field (a Gaussian beam), generated by: a 1-D periodic array of gold nanorods (green line), a hotspot of an individual gold trimer (red line), and a hot spot of a trimer located in between the array of nanorods – the two-scale structure (blue line). The hot spot is in the middle of a gap between two nanospheres of the trimer. The simulation is obtained assuming: the diameter of gold nanorods, the diameter of nanospheres, the gap spacing between nanospheres, and the period of the nanorod array are kept constant as 80nm, 40nm, 5nm, and 496 nm, respectively.

The capability of obtaining large field enhancement has been extensively utilized in SERS spectroscopy [41] to detect trace amount of biochemical analytes [42]. Here we demonstrate that our two-scale structure is suitable for this particular and important application. In Fig. 6 we consider a 1-D periodic set of gold nanorods on a glass substrate with the vacuum above. SERS involves at least two wavelengths in the measurement process, the incident field – here 785 nm – and the Raman scattered light which is reduced in energy – here we assume 850 nm for simplicity based on a 1000 cm$^{-1}$ vibrational frequency [15,43]. Thus, we design the period of our structure to attain the largest multiplicative enhancement at these two wavelengths. We use gold nanorods with square cross section with 150 nm side size and with period of 900 nm. We consider a gold oligomer placed on top of the substrate in the middle of the nanorod's array unit cell, where the diameter of each nanosphere is 80 nm and the gap spacing of them is 0.9 nm. The value for gap spacing is taken from [44,45,15] which shows reproducible gap spacings of 0.9 nm. Moreover, [15] shows that the majority of oligomers have a resonance wavelength near that of linear trimers, therefore, in our model we use a single linear trimer between the nanorod array arranged parallel to the nanorods. Full-wave simulations in Fig. 6 are based on the frequency domain finite element method (FEM), implemented in the commercial software CST Microwave Studio by Computer Simulation Technology AG. We consider normal plane wave excitation with electric field polarized along the nanorods and the linear trimer axis. Fig. 6 depicts the electric field enhancement achieved by individual trimer and the combination of trimer and gold nanorod array (i.e., by the proposed two-scale method).

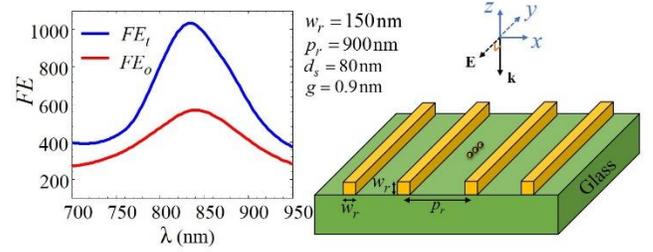

Fig. 6. Electric field enhancement versus wavelength of incident light generated by an individual gold trimer (red line) and by the combination of a periodic array of nanorods and trimer, i.e., the proposed two-scale structure (blue line). Field enhancement is calculated in the trimer hotspot. The structure is located on a glass substrate, with the vacuum above. The side width of gold nanorods with square cross section is 150 nm whereas the array period is 900 nm. The diameter of gold nanospheres and the gap between them is 80nm and 0.9nm, respectively. The structure is excited by a normal plane wave polarized along the nanorods and trimer axis.

To evaluate the performance of our two-scale structure, we define a figure of merit as

$$FoM = \left|FE_t/FE_o\right|^2_{\lambda=785} \times \left|FE_t/FE_o\right|^2_{\lambda=850} \qquad (13)$$

where $FE_t$ and $FE_o$ are the field enhancement of the two-scale structure and the individual oligomers at 785 nm and 850 nm, that are the incident and Raman scattered wavelengths, respectively. $FoM$ is our figure of merit that shows the enhancement of the SERS enhancement factor from incorporating our nanorod array, compared to the case without array of nanorods. According to Fig. 6, $\left|FE_t/FE_o\right|^2_{\lambda=785}$ is 2.31 and $\left|FE_t/FE_o\right|^2_{\lambda=850}$ is 3.06, yielding an $FoM$ of 7.07. These results theoretically demonstrate the benefit of our two-scale structure.

### C. Experimental Results

We now verify our theoretical predictions of augmented electric field enhancement due to the exploitation of Rayleigh anomaly through SERS measurements. Here SERS is generated by a surface fabricated using two completely different fabrication methods. As described in the methods section (C-D), we use a simple self-assembly method to deposit colloidal gold nanosphere oligomers on a glass substrate and fabricate periodic gold nanorods on a glass substrate through standard electron beam lithography. Once the rods are fabricated, gold nanospheres are chemically crosslinked onto the glass with an amine coupling. This process is repeated yielding oligomers. Previous work has shown that diffusion drives the formation of gold nanosphere oligomers suitable for investigation in this work [13]. Fig. 7 (a) and (b) depict scanning electron microscopy (SEM) images of gold nanosphere oligomers deposited on bare glass, and also deposited within the gold nanorod array, respectively. In Fig. 7 one may observe similar characteristics of gold nanosphere oligomers in both cases. The oligomers are separated by sufficient distance to avoid intra-oligomer strong coupling since in this paper we just want to show the oligomers-nanorods combined effect. Previous work has shown that the resonance frequency of close-packed gold nanosphere oligomer is nearly the same as the linear oligomers' with the same number of particles along the axis of polarization [15]. Consider a close-packed hexamer (as shown in Fig. 7 of [15].): three nanospheres are along any given axis of polarization. For such an arrangement of nanospheres, the oligomer resonates at nearly the same frequency as a linear trimer [15]. We have chosen 80 nm gold nanosphere oligomers because the resonance frequency of dimers and trimers of these sizes are best suited for providing large field enhancements at both 785 nm and approximately 850 nm.

Fig. 7(c) depicts the SERS spectra of standard SERS analyte benzenethiol (BZT) obtained using a surface with the two-scale structure and also (for comparison) using a surface covered only with colloidal oligomers, acquired as outlined in the methods section. Here, we use the same substrate for SERS measurements within (blue curve) and away from the gold nanorods (red curve), but on the same glass substrate, allows to minimize variation of oligomer enhancement and focus instead on the field enhancement augmentation due to the nanorod array (blue curve). Both spectra exhibit a large fluorescence peak from the glass microscope slide that appears in this background-subtracted data as a broad peak from 875 nm to 885 nm. Comparison of the two spectra at BZT's in-plane ring breathing mode at 1000 cm$^{-1}$ (observed at 852 nm) reveals a signal enhancement of 5.84 times for the surface made of the two-scale structure compared to the surface made of individual oligomers. This experimentally observed enhancement is in good agreement with the predicted FOM for this system and demonstrates that the Rayleigh anomaly field enhancement is relatively robust to scattering caused by fabrication defects.

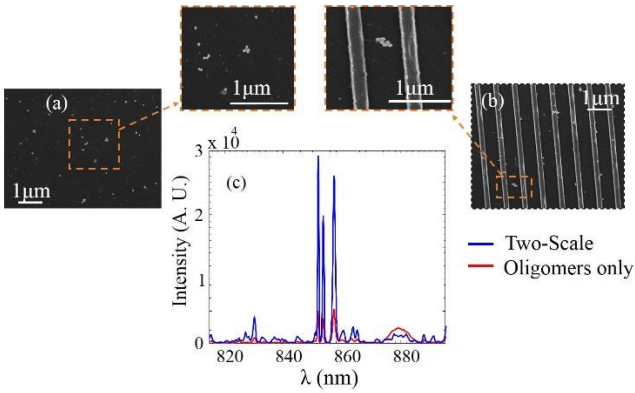

Fig 7. Scanning Electron Microscopy (SEM) images of gold nanospheres oligomer without nanorod array(a) and when placed within the gold nanorod array (b). In both cases, the structure is deposited on glass. The gold nanorods have a period of 900 nm and square cross section with 150nm side. The gold nanosphere oligomers are expected to have an 80 nm diameter and a 0.9 nm gap from nanosphere to nanosphere. (c) shows the SERS spectra relative to the proposed two-scale structure shown in Fig. 7(b) and to the oligomers-only structure shown in Fig. 7(a), versus wavelength. The two-scale structure provides much stronger SERS signal.

## 5. Conclusion

By combining plasmonic resonance of metallic colloidal oligomers with a Rayleigh anomaly (i.e., a structural field enhancement due to coherent superposition of fields) of a 1-D periodic array of nanorods, we overcame the limitation for field enhancement imposed by losses and nonlocality of the gold nanospheres' dielectric response. The method is called "two-scale" because it involves two completely different and complementary fabrication methods for enhancing electric fields: direct colloidal assembly and lithographic techniques. Specifically, the field enhancement of our proposed two-scale structure has been proved superior to the field enhancement of a surface made of colloidal oligomers only (i.e, without the 1-D periodic array of nanorods). We have provided a model that explains in physical and mathematical terms such field enhancement and provided an experimental demonstration of the proposed structure showing augmented SERS spectra of BZT analytes. We expect these two-scale fabricated substrates to find several applications in medical diagnostics, solar cells, sensors, and single molecule detectors.

## Methods

### A. The Magnetic Vector Potential Green's Function

The spectral representation of the magnetic vector potential Green's function $G^{\infty}(\mathbf{r},\mathbf{r}_0)$ for a periodic array of line sources along the $y$-direction where $\mathbf{r}$ is the observation point and $\mathbf{r}_0$ is the reference source point (the source of a reference rod) is defined as [46]

$$G^{\infty}(\mathbf{r},\mathbf{r}_0) = \sum_{p=-\infty}^{\infty} \frac{i}{p_r} \frac{e^{ik_{xp}(x-x_0)}}{2k_{zp}} e^{ik_{zp}|z-z_0|} \quad (14)$$

where $k_{zp} = \sqrt{k_h^2 - k_{xp}^2}$ and $k_{xp} = k_x + 2\pi p/p_r$ is the $p^{\text{th}}$ Floquet harmonic wavenumber. Here $k_x p_r$ is the phase shift from one line source to the other, imposed by the incident field. Note that here, $G^{\infty}$ is the Green's function of the magnetic vector potential $A$ (both are scalar and in the $y$ direction) and is related to the $y$ component of the electric field as $E(\mathbf{r}) = i\omega\mu_0 A(\mathbf{r})$. We apply the Ewald method to accelerate the convergence of the above series, and to have explicit both spatial and spectral singularities. Therefore, the Green's function is written as the summation of the spatial and the spectral parts as

$$G^{\infty}(\mathbf{r},\mathbf{r}_0) = \sum_{n=-\infty}^{\infty} g_n^E(\mathbf{r},\mathbf{r}_0) + \sum_{p=-\infty}^{\infty} \tilde{g}_p^E(\mathbf{r},\mathbf{r}_0) \quad (15)$$

where, the first term is the modified spatial representation and the second term is the spectral representation of the Ewald formulation which are presented in Eq. (5) and Eq. (6) of [47] respectively. The spatial part contains the logarithmic singularity in $g_0^E$ when $\mathbf{r} \to \mathbf{r}_0$ whereas the spectral part contains the spectral singularities when $k_{zp} \to 0$. The "regularized" Green's function in Eq. (6) is defined as $\breve{G}^{\infty}(\mathbf{r}_0,\mathbf{r}_0) = \lim_{\mathbf{r} \to \mathbf{r}_0} [G^{\infty}(\mathbf{r},\mathbf{r}_0) - G(\mathbf{r},\mathbf{r}_0)]$ where $G(\mathbf{r},\mathbf{r}_0) = (i/4)H_0^{(1)}(k_h|\mathbf{r}-\mathbf{r}_0|)$, represents the scalar 2-D Green's function of the magnetic vector potential of the array, evaluated at the reference nanorod location $\mathbf{r}_0$, without considering the field contribution of the same nanorod at $\mathbf{r}_0$. Also, this regularized Green's function is represented based on the Ewald method as

$$\breve{G}^{\infty}(\mathbf{r}_0,\mathbf{r}_0) = \sum_{p=-\infty}^{\infty} \frac{i}{2p_r k_{zp}} \text{erfc}\left(\frac{-ik_{zp}}{2E}\right) +$$
$$\sum_{\substack{n=-\infty \\ n \neq 0}}^{\infty} \frac{1}{4\pi} e^{ik_x n p_r} \left(\sum_{q=0}^{\infty} \left(\frac{k_h}{2E}\right)^{2q} \frac{1}{q!} E_{q+1}(R_n^2 E^2)\right) + \quad (16)$$
$$\frac{1}{2\pi} \ln\left(\frac{k_h}{E}\right)$$

where $R_n = |n|p_r$ and $E$ is the splitting parameter and the optimum value to minimize the overall number of terms needed to calculate the Green's function is $E_{opt} = \sqrt{\pi}/p_r$ [47], though the spectral singularity remains and it plays a major role in the field enhancement.

### B. Limit of Field Enhancement at Rayleigh anomaly

We prove here that in our periodic structure, when the wavelength of the incident light matches $\lambda_r$ in Eq. (12), the field enhancement due to the nanorods in the middle of two adjacent nanorods ($FE_r$) is exactly 2. Indeed, we note that at that wavelength, $|k_{xp}| = |k_x + 2\pi p / p_r| = k_h$, hence $k_{zp} = 0$. Therefore, $G^\infty(\mathbf{r}, \mathbf{r}_0)$ in Eq. (14) and Eq. (15) tends to infinity. On the other hand, also the first summation in Eq. (16) in calculating $\breve{G}^\infty(\mathbf{r}_0, \mathbf{r}_0)$ is singular, i.e., $\breve{G}^\infty(\mathbf{r}_0, \mathbf{r}_0)$ tends to infinity as well. Assuming the field observation point is $\mathbf{r} = \lambda_r / 2 \, \hat{\mathbf{x}}$, i.e., exactly in the middle of two nanorods, and $\mathbf{r}_0 = (0,0)$, by substituting Eq. (7) in Eq. (5), we get

$$E_r^{\text{ext}}(\mathbf{r}) = E^{\text{inc}}(\mathbf{r}) + i\omega\mu_0 \frac{\alpha_r E^{\text{inc}}(\mathbf{r}_0)}{1 - i\omega\mu_0 \alpha_r \breve{G}^\infty(\mathbf{r}_0, \mathbf{r}_0)} G^\infty(\mathbf{r}, \mathbf{r}_0) \quad (17)$$

Note that when $k_{zp} \to 0$ both the numerator and denominator of the second term in Eq. (17) tend to infinity because of the singularity in both $G^\infty(\mathbf{r}, \mathbf{r}_0)$ and $\breve{G}^\infty(\mathbf{r}_0, \mathbf{r}_0)$. Therefore, by taking the limit of $E_r^{\text{ext}}(\mathbf{r})$, we get

$$\lim_{k_{zp} \to 0} E_r^{\text{ext}}(\mathbf{r}) = E^{\text{inc}}(\mathbf{r})\left[1 - e^{-i\mathbf{k}_h \cdot (\mathbf{r} - \mathbf{r}_0)}\right] = 2E^{\text{inc}}(\mathbf{r}) \quad (18)$$

which proves that the field enhancement due to the Rayleigh anomaly is equal to 2.

### C. Fabrication of nanorod array substrate

Array of nanorods with diameter of 150 nm is fabricated using standard negative tone electron beam lithography. Briefly, glass coated with 50nm gold and a 5nm Titanium adhesion layer (Ted Pella) substrates are cleaned and then Ma-N 2401 (Microchem) is spin coated for use as the photoresist. The resist is exposed in a Magellan XHR SEM (FEI) at 25 pA and 30kV and subsequently developed in Ma-D (Microchem). The substrate is then etched with ion milling (IntIvac) to remove gold in the unexposed regions leaving nanorods. The titanium layer is etched with a 1:4 $H_2O_2$:$H_2SO_4$ solution to ensure that nanoparticle attachment will occur on the plane of the array, and to remove remaining photoresist.

### D. Nanoparticles attachment

Nanorod array substrates are submerged in 0.5mMol (3-Aminopropyl) triethoxysilane (APTES) (Sigma Aldrich) in deionized water overnight to form a self-assembled monolayer selectively on glass yielding primary amine groups. 80 nm lipoic acid functionalized gold nanoparticles synthesized via a seeded growth method described by [48]. These particles are functionalized with carboxylic acid end groups by replacing the solution with DI water adjusted to pH 11.67 with sodium hydroxide (Sigma) and 0.1 mMol Lipoic acid and allowing the particles to sit overnight. The particles are washed with DI water and are deposited onto the glass surface by carbodiimide crosslinking with the APTES monolayer as described elsewhere. Diffusion drives the formation of random nanoparticle oligomers [14].

### E. Surface-enhanced Raman scattering spectroscopy measurements

Nanorod array substrates with attached Au nanosphere oligomers – two-scale structures - are soaked overnight in 0.5 mMol benzenethiol in ethanol solution and then rinsed thoroughly prior to all SERS measurements. SERS measurements of two-scale structures, and Au nanosphere oligomers on glass are acquired in a Renishaw inVia Raman microscope for 10s at 0.125 mW using a 50x air objective lens. Spectra are background subtracted and Savitzky Golay smoothed. Spectra are obtained on and off of nanorod arrays using the same substrate to reduce possible sample to sample variation.

**Acknowledgment**. Authors acknowledge support from the National Science Foundation, NSF-SNM-1449397. We are grateful to CST Simulation Technology AG for letting us use the simulation tool CST Microwave Studio that was instrumental in this analysis. The authors would like to thank Dr. Caner Guclu for fruitful discussions on the Ewald's method.